\documentclass[11pt]{article}

\usepackage[T1]{fontenc}
\usepackage{chao}
\usepackage{mathtools}
\usepackage[capitalise,noabbrev]{cleveref}
\usepackage{etoolbox}
\crefname{theorem}{Theorem}{Theorems}
\crefname{lemma}{Lemma}{Lemmas}
\crefname{corollary}{Corollary}{Corollaries}
\crefname{proposition}{Proposition}{Propositions}
\crefname{conjecture}{Conjecture}{Conjectures}
\crefname{definition}{Definition}{Definitions}
\crefname{problem}{Problem}{Problems}
\crefname{example}{Example}{Examples}
\crefname{remark}{Remark}{Remarks}
\crefname{appendix}{Appendix}{Appendices}
\apptocmd{\appendix}{\crefalias{section}{appendix}}{}{}
\usepackage{subcaption}
\usepackage{array}
\title{All minimum cuts of symmetric submodular functions}
\author{Chao Xu}
\date{}
\hypersetup{pdftitle={All minimum cuts of symmetric submodular functions},pdfauthor={Chao Xu}}
\AtBeginEnvironment{thebibliography}{\small}

\begin{document}

\maketitle

\begin{abstract}
We compute Cunningham's canonical decomposition of a symmetric submodular function on \(n\) elements in \(O(n^3)\) time. The decomposition represents all minimum cuts in linear space. For the pair \(s,t\) produced by Queyranne's algorithm, we construct in \(O(n^2)\) time a digraph representing exactly the minimum \(st\)-cuts of the function. Using this construction in Chekuri and Xu's decomposition algorithm gives the cubic bound.
\end{abstract}

\section{Introduction}\label{sec:introduction}

Let \(f:2^V\to\mathbb R\) be a symmetric submodular function on \(n\) elements, given by a value oracle. Its minimum cuts are the bipartitions of \(V\) into nonempty parts minimizing the value of either part.

Queyranne \cite{queyranne1998} finds a minimum cut in \(O(n^3)\) time. Nagamochi and Ibaraki \cite{nagamochi1998} find all inclusionwise minimal nonempty proper minimizers using \(O(n^3)\) value-oracle calls. Cunningham's canonical decomposition \cite{cunningham1983,cunninghamEdmonds1980} represents all minimum cuts. By finding all minimum cuts, we mean finding the canonical decomposition.
\begin{theorem}\label{thm:main}
Given a symmetric submodular \(f:2^V\to\mathbb R\) on \(n\) elements, all of its minimum cuts can be found in \(O(n^3)\) time.
\end{theorem}

The vertex order produced by Queyranne's algorithm is called a \emph{Queyranne ordering}. A minimum \(st\)-cut minimizes the value among cuts separating \(s\) and \(t\). For the final vertices \(s,t\) of the ordering, a digraph representation encodes the \(s\)-sides of all minimum \(st\)-cuts as vertex sets containing \(s\) and excluding \(t\) with no outgoing arc.

By following Chekuri and Xu's algorithm \cite[Section~4.1]{chekuri2018}, one observes that
if the digraph representation can be constructed in time \(T\), then the canonical decomposition of a symmetric submodular function of a \(n\)-element can be constructed in \(O(n(T+n^2))\) time.

Arikati and Mehlhorn \cite{arikati1999} construct a maximum flow for the final pair of a graph's maximum adjacency ordering. Chekuri and Xu \cite{chekuri2018} do so for a hypergraph's tight ordering in an associated digraph. In both constructions, the residual graphs represent all minimum \(st\)-cuts. We show such a representation for symmetric submodular functions also exists and can be found in $T=O(n^2)$ time.

\begin{theorem}\label{thm:flow}
Given a value oracle for a symmetric submodular \(f:2^V\to\mathbb R\) and a Queyranne ordering \(v_1,\ldots,v_n\), a digraph representation of all minimum \(st\)-cuts, where \(s=v_{n-1}\) and \(t=v_n\), can be constructed in \(O(n^2)\) time.
\end{theorem}

Hence \cref{thm:main} holds when the above theorem holds, and the rest of the note is about proving \cref{thm:flow}.

\section{Preliminaries}\label{sec:preliminaries}

Write \(A+v=A\cup\{v\}\), \(A-v=A\setminus\{v\}\), and \(f(v)=f(\{v\})\). For a vector \(x\), write \(x(A)=\sum_{u\in A}x(u)\). A function \(f:2^V\to\mathbb R\) is \emph{symmetric} if \(f(A)=f(V\setminus A)\) and \emph{submodular} if \(f(A)+f(B)\ge f(A\cup B)+f(A\cap B)\) for all \(A,B\subseteq V\). For submodular \(f\), its minimizers over \(2^V\) are closed under union and intersection.

For symmetric \(f\), a \emph{cut} is a bipartition \((S,T)\) into nonempty parts, of value \(f(S)\). It is an \emph{\(st\)-cut} if \(s\in S\) and \(t\in T\). A minimum cut minimizes the value over all cuts, and a minimum \(st\)-cut does so among \(st\)-cuts. A set is \emph{closed} in a digraph if no arc leaves it. A digraph represents all minimum \(st\)-cuts if its closed sets containing \(s\) and excluding \(t\) are exactly the \(s\)-sides of minimum \(st\)-cuts.

For a vertex ordering $v_1,\ldots,v_n$, we write \(V_j=\{v_1,\ldots,v_j\}\), with \(V_0=\varnothing\). A \emph{Queyranne ordering} \cite{chekuri2018} chooses each \(v_{j+1}\) to minimize \(f(V_j+v)-f(v)\) over \(v\notin V_j\). It takes \(O(n^2)\) time to compute. The final pair \(s=v_{n-1},t=v_n\) is \emph{pendent}: \((V-t,\{t\})\) is a minimum \(st\)-cut \cite{queyranne1998}.

For submodular \(f:2^V\to\mathbb R\) with \(f(\varnothing)=0\), a \emph{base} is a vector \(x\in\mathbb R^V\) with \(x(A)\le f(A)\) for all \(A\subseteq V\) and equality on \(V\). A set is \emph{tight} when equality holds. The \emph{greedy base} of an ordering $v_1,\ldots,v_n$ is the base defined as \(x(v_i)=f(V_{i})-f(V_{i-1})\). For each $i$, $V_i$ is tight for $x$ \cite{edmonds1970}.

Given a greedy base and its ordering, a digraph whose closed sets are exactly the tight sets can be constructed in \(O(|V|^2)\) time  \cite{bixby1985}\cite[Algorithm~5.2.1 and Theorem~5.2.11]{topkis1998}.

Vertex balance is incoming minus outgoing flow. We use complementary slackness \cite[Section~4]{kolmogorov2012}.

\begin{lemma}\label{lem:minimizer-representation}
Let \(r:2^U\to\mathbb R_{\ge0}\) be submodular with \(r(\varnothing)=r(U)=0\), and let \(x\) be a base of \(r\). Suppose a nonnegative flow on a digraph on \(U\) has balance \(x\), and no flow leaves any minimizer of \(r\). Adding the reversals of its positive-flow arcs to a digraph whose closed sets are the \(x\)-tight sets gives a digraph whose closed sets are exactly the minimizers of \(r\).
\end{lemma}

\begin{proof}
A set is closed in the augmented digraph exactly when it is tight and no positive flow enters it. If \(B\) minimizes \(r\), no flow leaves \(B\), so \(x(B)\) is its incoming flow. Since \(x\) is a base, \(0\le x(B)\le r(B)=0\), establishing both conditions. Conversely, if \(B\) is tight and no flow enters it, \(x(B)\) is minus its outgoing flow, so \(0\le r(B)=x(B)\le0\).
\end{proof}

\section{The algorithm and proof}\label{sec:certificate}

Fix a Queyranne ordering \(v_1,\ldots,v_n\) of \(f\). Set \(w(v_j)=(f(V_{j+1})-f(v_{j+1}))-(f(V_j)-f(v_j))\) for \(1\le j<n\).
For \(1\le i\le n\), we define the \emph{slack} \(r_i:2^{V_{i-1}}\to\mathbb R\) by \(r_i(A)=f(A+v_i)-f(v_i)-w(A)\).

We refine Queyranne's pendent-pair induction \cite{queyranne1994} and show that each \(r_i\) is nonnegative.

\begin{lemma}\label{lem:maximum-index}
For a submodular \(f\) and a Queyranne ordering, each \(r_i\) is nonnegative. If \(r_i(A)=0\) for nonempty \(A\subseteq V_{i-1}\) and \(v_j\) has largest index in \(A\), then \(r_i(V_j)=0\) and \(r_j(A-v_j)=0\).
\end{lemma}

\begin{proof}
Telescoping gives \(w(V_{i-1})=f(V_i)-f(v_i)\), and the ordering rule gives \(r_i(V_j)\ge0\) for \(j<i\). Induct on \(i\). For nonempty \(A\subseteq V_{i-1}\), let \(v_j\) have largest index in \(A\). Then
\[
\begin{aligned}
r_i(A)={}&r_j(A-v_j)+r_i(V_j)\\
&+f(A+v_i)-f(A)-f(V_j+v_i)+f(V_j).
\end{aligned}
\]
The first term is nonnegative by induction, the second by the ordering rule, and the last line by submodularity. If \(r_i(A)=0\), then $r_j(A-v_j)=r_i(V_j)=0$.
\end{proof}

By \cref{lem:maximum-index}, \(w(A)\) is an additive lower bound on the cost \(f(A+v_i)-f(v_i)\) of adding \(A\) when \(v_i\) is present. Telescoping makes this bound exact on \(V_{i-1}\).

For \(A\subseteq V_{n-2}\), symmetry and \(w(V_{n-2})=f(t)-f(s)\) give
\begin{equation}\label{eq:pair-identity}
r_n(A)+r_{n-1}(V_{n-2}\setminus A)=2\bigl(f(A+t)-f(t)\bigr).
\end{equation}
Hence \(A+t\) is the \(t\)-side of a minimum \(st\)-cut exactly when $r_n(A)=r_{n-1}(V_{n-2}\setminus A)=0$.

Each \(r_i\) is nonnegative by \cref{lem:maximum-index} and submodular because \(w\) is additive, with \(r_i(\varnothing)=r_i(V_{i-1})=0\) by definition and telescoping. To apply \cref{lem:minimizer-representation} and obtain digraphs representing minimizers, we construct greedy bases of \(r_n,r_{n-1}\) and nonnegative flows whose vertex balances equal these bases, with no flow leaving a minimizer. The flows come from forests defined by prefixes \(V_j\) with \(r_i(V_j)=0\).

For \(1\le i\le n\), put \(J_i=\{j:0\le j<i,\ r_i(V_j)=0\}\), which contains \(0,i-1\). Consecutive \(k<j\) in \(J_i\) partition \(V_{i-1}\) into intervals \(V_j\setminus V_k\). In each interval \(V_j\setminus V_k\), remove its rightmost vertex \(v_j\) and recursively partition the remainder into child intervals at consecutive boundaries in \(\{k\}\cup\{\ell\in J_j:\ell>k\}\), stopping at singletons. For each fixed \(i\), the resulting intervals are nested or disjoint.

Encode this interval hierarchy as a rooted forest \(F_i\) on \(V_{i-1}\): label intervals by their right endpoints and direct child intervals to their parents. Identify each subtree with its interval, and let \(F_1\) be empty. For a forest \(F\), write \(F[U]\) for the induced forest on \(U\). Deleting a prefix \(V_k\) from these forests preserves the parents of remaining vertices and truncates their subtree intervals. Hence the tree on \(V_j\setminus V_k\) is obtained by attaching the roots of \(F_j[V_{j-1}\setminus V_k]\) to \(v_j\).

Let \(\pi_i\) order \(V_{i-1}\) in forest preorder, visiting each interval's right endpoint before its children and processing disjoint intervals from left to right. We compute the greedy base \(x_i\) of \(r_i\) in the order \(\pi_i\). Since \(x_i\) is a base, \(x_i(V_{i-1})=r_i(V_{i-1})=0\).

\begin{lemma}\label{lem:forest}
For a submodular \(f\) and a Queyranne ordering, every minimizer of \(r_i\) is closed in \(F_i\). Every subtree \(T\) of \(F_i\) satisfies \(x_i(T)\le0\).
\end{lemma}

\begin{proof}
For closure, induct on \(i\). Let \(A\) minimize \(r_i\), and take consecutive \(k<j\) in \(J_i\). Submodularity makes \(A\cap V_j\) a minimizer. If it meets \(V_j\setminus V_k\), its largest index belongs to \(J_i\) by \cref{lem:maximum-index}, so it is \(j\). The same lemma gives \(r_j(A\cap V_{j-1})=0\). Induction gives closure under the inherited edges of \(F_j\), and \(v_j\in A\) gives closure under the new edges.

Let \(T=V_j\setminus V_k\) be a subtree rooted at \(v_j\). If \(v_a\) is the parent of \(v_j\), taking \(a=i\) at a root, then \(j\in J_a\) by construction. Earlier trees and left sibling subtrees along the ancestor path partition \(V_k\), so the vertices preceding \(T\) in \(\pi_i\) are \(V_k\) and the strict ancestors of \(v_j\). Let \(X\) consist of those ancestors and \(v_i\). Then \(v_a\in X\) and \(X\cap V_j=\varnothing\). Summing the differences defining \(x_i\) over \(T\) and using submodularity gives
\[
\begin{aligned}
x_i(T)
&=f(V_j\cup X)-f(V_k\cup X)-w(T)\\
&\le f(V_j+v_a)-f(V_k+v_a)-w(T)\\
&=r_a(V_j)-r_a(V_k)=-r_a(V_k)\le0.
\end{aligned}
\]
\aftermath
\end{proof}

In each \(F_i\), write \(T_u\) for the subtree vertex set at \(u\) and \(p(u)\) for its parent at a nonroot.

\begin{figure}[ht]
\centering
{\small
\begin{algorithm}
\textsc{AllMinSTCuts}\((f,v_1,\ldots,v_n)\)\\
Compute \(w\) and \(J_1,\ldots,J_n\) from the values \(f(V_j+v_i)\), \(0\le j<i\le n\).\\
\(F_1\gets\varnothing\).\\
for \(i=2,\ldots,n\)\\
\>\(F_i\gets\) the forest of isolated vertices \(V_{i-1}\).\\
\>for consecutive \(k<j\) in \(J_i\)\\
\>\>Copy \(F_j[V_{j-1}\setminus V_k]\) into \(F_i\), attaching its roots to \(v_j\).\\
for \(i\in\{n-1,n\}\)\\
\>\(\pi_i\gets\) the preorder of \(F_i\), and \(x_i\gets\) the greedy base of \(r_i\) along \(\pi_i\).\\
\>\(D_i\gets\) a digraph on \(V_{i-1}\) whose closed sets are the tight sets of \(x_i\).\\
\>Add \(p(u)\to u\) to \(D_i\) for each nonroot \(u\) with \(x_i(T_u)<0\).\\
return the digraph \(D\) on \(V\) formed by \(D_{n-1}\) and the reversal of \(D_n\).
\end{algorithm}\par}
\caption{Constructing a digraph representation of all minimum \(st\)-cuts.}
\label{fig:pair-algorithm}
\end{figure}

\begin{proof}[\cref{thm:flow}]
Fix \(i\in\{n-1,n\}\). On the forest \(F_i\), assign flow \(-x_i(T_u)\) to each arc \(u\to p(u)\). These flows are nonnegative by \cref{lem:forest}. The sums \(x_i(T)\) over the trees \(T\) of \(F_i\) are nonpositive and total zero, so each is zero. At a nonroot \(u\), incoming flow is \(x_i(u)-x_i(T_u)\), while outgoing flow is \(-x_i(T_u)\), giving balance \(x_i(u)\). At a root, outgoing flow is zero and \(x_i(T_u)=0\), so the same balance holds. Every minimizer of \(r_i\) is closed in \(F_i\), so no flow leaves it. Hence \cref{lem:minimizer-representation} shows that \(D_i\) in \cref{fig:pair-algorithm} represents the minimizers of \(r_i\).

For \(s\in S\) and \(t\notin S\), put \(A=V_{n-1}\setminus S\subseteq V_{n-2}\). By arc reversal, \(S\) is closed in \(D\) exactly when \(r_n(A)=r_{n-1}(V_{n-2}\setminus A)=0\), which characterizes minimum \(st\)-cuts by \cref{eq:pair-identity}.

The values \(f(V_j+v_i)\), \(0\le j<i\le n\), cost \(O(n^2)\) oracle calls. Store \(F_i\) in a parent array, with \(i\) in each root's entry. For consecutive \(k<j\) in \(J_i\), copy entries \(k+1,\ldots,j-1\) from \(F_j\) and set the entry for \(v_j\) to \(i\). For each copied root, the stored value \(j\) records its new parent \(v_j\). Each entry is processed a constant number of times, giving \(O(i)\) time for \(F_i\). Both preorders, bases, and subtree sums take \(O(n)\) further work and calls. Each tightness test on a set \(B\) costs one call to \(f\) and constant work, maintaining \(w(B)\) and \(x_i(B)\) under single-element updates. The tight-set construction in \cref{sec:preliminaries} gives \(O(n^2)\) total time and oracle calls.
\end{proof}

\paragraph{Acknowledgments}

Chao thanks Satoru Fujishige and Yutaro Yamaguchi for discussions about extending the result to symmetric submodular functions. At the time, Chao gave Yamaguchi the impression that he thought he had a proof and that the result would follow soon. That optimism was premature: the solution came more than ten years later. Finally, he thanks Kyriakos Axiotis and Dimitris Fotakis for early discussions of this problem.

\paragraph{AI disclosure}
\cref{thm:flow} was resolved using GPT-6 Astra with Xean in 13 hours~\cite{xean}. Opus 5.5 was also used to edit the manuscript and review the arguments. The author checked the proof and takes responsibility for the final manuscript.

\bibliographystyle{ieeetr}
\bibliography{references}

\begin{thebibliography}{10}

\bibitem{queyranne1998}
M.~Queyranne, ``Minimizing symmetric submodular functions,'' {\em Mathematical Programming}, vol.~82, no.~1--2, pp.~3--12, 1998.

\bibitem{nagamochi1998}
H.~Nagamochi and T.~Ibaraki, ``A note on minimizing submodular functions,'' {\em Information Processing Letters}, vol.~67, no.~5, pp.~239--244, 1998.

\bibitem{cunningham1983}
W.~H. Cunningham, ``Decomposition of submodular functions,'' {\em Combinatorica}, vol.~3, no.~1, pp.~53--68, 1983.

\bibitem{cunninghamEdmonds1980}
W.~H. Cunningham and J.~Edmonds, ``A combinatorial decomposition theory,'' {\em Canadian Journal of Mathematics}, vol.~32, no.~3, pp.~734--765, 1980.

\bibitem{chekuri2018}
C.~Chekuri and C.~Xu, ``Minimum cuts and sparsification in hypergraphs,'' {\em SIAM Journal on Computing}, vol.~47, no.~6, pp.~2118--2156, 2018.

\bibitem{arikati1999}
S.~R. Arikati and K.~Mehlhorn, ``A correctness certificate for the {Stoer--Wagner} min-cut algorithm,'' {\em Information Processing Letters}, vol.~70, no.~5, pp.~251--254, 1999.

\bibitem{edmonds1970}
J.~Edmonds, ``Submodular functions, matroids, and certain polyhedra,'' in {\em Combinatorial Structures and Their Applications}, pp.~69--87, New York: {Gordon and Breach}, 1970.

\bibitem{bixby1985}
R.~E. Bixby, W.~H. Cunningham, and D.~M. Topkis, ``The partial order of a polymatroid extreme point,'' {\em Mathematics of Operations Research}, vol.~10, no.~3, pp.~367--378, 1985.

\bibitem{topkis1998}
D.~M. Topkis, {\em Supermodularity and Complementarity}.
\newblock Princeton University Press, 1998.

\bibitem{kolmogorov2012}
V.~Kolmogorov, ``Minimizing a sum of submodular functions,'' {\em Discrete Applied Mathematics}, vol.~160, no.~15, pp.~2246--2258, 2012.

\bibitem{queyranne1994}
M.~Queyranne, ``A combinatorial algorithm for minimizing symmetric submodular functions,'' Tech. Rep. 393, Technische Universit\"at Berlin, 1994.

\bibitem{xean}
C.~Xu, ``{Xean}.'' \url{https://github.com/chaoxu/xean}.

\end{thebibliography}

\end{document}